\documentclass[twocolumn,aps,prl,floatfix]{revtex4}
\usepackage{graphicx,amsmath,subfigure,epsfig,psfrag}
\usepackage{verbatim}
\input epsf

\def\be{\begin{equation}}
\def\ee{\end{equation}}
\def\ba{\begin{eqnarray}}
\def\ea{\end{eqnarray}}

\makeatother
\begin{document}

\title{``Exact'' Algorithm for Random-Bond Ising Models in 2D}
\author{Y.~L.~Loh and E.~W.~Carlson}
\affiliation{Department of Physics, Purdue University, West Lafayette, IN  47907 }
\date{\today}

\begin{abstract}
We present an efficient algorithm for calculating the properties of Ising models in two dimensions,
directly in the spin basis, without the need for mapping to fermion or dimer models.  
The algorithm gives numerically exact results for the partition function 
and correlation functions at a single temperature on any planar network of $N$ Ising spins in $O(N^{3/2})$ time or less.
The method can handle continuous or discrete bond disorder and is especially efficient in the case of bond or site dilution, where it executes in $O(L^2{\rm ln}L)$ time near the percolation threshold.
We demonstrate its feasibility on the ferromagnetic Ising model and the $\pm J$ random-bond Ising model (RBIM) 
and discuss the regime of applicability in cases of full frustration such as the Ising antiferromagnet on a triangular lattice.
\end{abstract}
\maketitle

%==============================================================================`
%\mysection{What the algorithm can do}
Ising models are the prototype system for studying phase transitions, critical phenomena, and disordered systems.  We present here an algorithm for computing the partition function and correlation functions in a class of 2D Ising models which is exact to machine precision
%EC ``numerically exact'',
%[[[[[i.e., its accuracy is limited by the precision of the floating-point numbers used in the computation; it suffers from roundoff error but not statistical error like Monte Carlo methods]]]]]]
and which works for any planar network of Ising spins with arbitrary bond strengths but without applied fields in the bulk.  Applications include random-bond Ising models (RBIM) (including $\pm J$ disorder, Gaussian disorder, site dilution, and bond dilution) and geometric frustration as in the case of triangular Ising antiferromagnets.

%==============================================================================`
%\mysection{Summary of how it works; efficiency}
Our algorithm is an extension of a bond propagation algorithm\cite{lobb} originally developed for resistor networks.  
The method works by successively integrating in and then integrating out spin degrees of freedom 
in a way that only introduces local changes to the network, 
in order to progressively move degrees of freedom to an open edge of the network, where they are eliminated.
%It is transparent to understand and derive, as 
It operates directly on the original spin network, \emph{without mapping to fermions or dimers} and requires negligible memory in addition to the $O(L^2)$ memory required to store the Ising bond strengths that define the problem.
The algorithm requires $O(L^3)$ time to compute the partition function $Z(T)$ of an $L\times L$ square lattice at a single temperature $T$; for bond-diluted problems near the percolation threshold it requires only $O(L^2 {\rm ln} L)$ time, which is the fastest method to our knowledge in this case.
\footnote{Ref.~\onlinecite{lobb} showed that the bond-propagation algorithm
requires only $O(L^2 ln L)$ time for a dilute resistor network near
percolation.  Since the number of bond propagations depends only on
the "network topology", we conclude that the computational time will scale the same way in the Ising case.}  
In comparison, the fermion network method takes $O(L^4)$ time\cite{merz}, spin-basis transfer matrix methods\cite{morgenstern1979} take $O(2^L)$ time, and the Pfaffian method with nested dissection takes $O(L^3)$ time\cite{chineseremainder,integer}.
%EC -- I merely moved to bib, to avoid [4]. [13] appearing.  
%\footnote{ 
%There are also methods\cite{saul1993,saul1994,chineseremainder} that require $O(L^5)$ time or more to compute the exact integer values of the $O(L^2)$ series coefficients of $Z(T)$, and are restricted to RBIMs with integer couplings.}
While our algorithm has superior speed to that
of Ref.~\cite{chineseremainder} only near the percolation threshold, we believe
there are advantages to a transparent algorithm which
operates directly in the spin representation.
In addition, our method is highly parallelizable, and can execute in as little as $O(L)$ time if a sufficient number of nodes are available. 
%How fast canGLV be with parallelization? O(L^5), but presumably that's getting all coefficients with integer accuracy.  So perhaps if they dropped the integer accuracy and calculated Z at one temp, they'd get parallelized O(L) like us.  

%%%%%%%% FIGURE:  Bond Propagation %%%%%%%%%%%%%%%
\begin{figure}[Htb]
{\centering
\subfigure[Lattice reduction]
{\resizebox*{0.9\columnwidth}{!}{\includegraphics{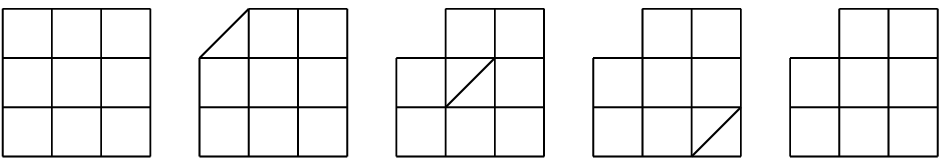}\label{fig:lattice}}}
\subfigure[A single bond propagation move]
{\resizebox*{0.9\columnwidth}{!}{\includegraphics{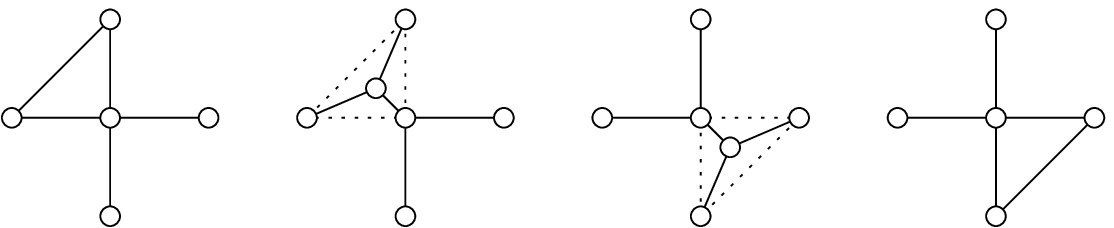}\label{fig:bond}}}
\par}
\caption{The bond propagation algorithm.\cite{lobb}  (a) 
Starting from one corner, the two outer bonds may be combined using a series reduction to make a single  diagonal bond.  
Then, the lattice can be reduced by successively using the bond propagation algorithm to move the diagonal bond out of the lattice.  Repeated applications of the algorithm reduce
the lattice to a single bond, corresponding to the effective resistance of the entire resistor network, or in the Ising case,  
corresponding to a reduced $2$-spin system with effective coupling $J_{\rm eff}$ whose partition function is equal to that of the original lattice.  
(b) A single bond propagation step, in which a $\Delta$-$Y$ 
and then a $Y$-$\Delta$ transformation are used to propagate one diagonal bond through
a $4$-fold coordinated node in the lattice.
}
\label{fig:algorithm}
\end{figure}
%%%%%%%%%%%%%%%%%%%%%%%%%%%%%%%%%%%%%%

%==============================================================================
%\mysection{Bond propagation for resistor networks}
%==============================================================================
%We first describe the original bond propagation algorithm as applied to resistor networks\cite{lobb}  and then develop the extension to the Ising system.   
We begin by describing the original bond propagation algorithm invented by Frank and Lobb \cite{lobb}.
The effective resistance of any 2D resistor network can be calculated swiftly and accurately by this algorithm.  
There are two basic transformations required:    a series reduction and the so-called $Y$-$\Delta$ transformation (along with its corresponding inverse).  
Using these ingredients, a 2D resistor network can be efficiently reduced to a single net resistance in the following way: Starting from the upper left corner in Fig.~\ref{fig:lattice},
use a series reduction to convert the corner into a diagonal bond.  Using the $Y$-$\Delta$ and $\Delta$-$Y$ transformations, this diagonal bond can be successively propagated diagonally down and to the right until it annihilates at an edge with open boundary conditions.  The way that one ``bond propagation" move is completed is illustrated in Fig.~\ref{fig:bond}.
First, the upper left $\Delta$ in Fig.~\ref{fig:bond} is converted into a $Y$.  This introduces one new node into the system.  The new node is now effectively shifted, in order to replace the node directly to its lower right, a ``move" which does not change the topology of the network.  Finally, the lower right $Y$ is converted into a $\Delta$.  In this way, a diagonal bond in any  lattice with coordination number $z \le 4$ can be ``propagated" diagonally.  Repeated bond propagation moves reduce the network to a single string of 
resistors in series, which is easily reducible to one effective resistor.  

%==============================================================================`
%\mysection{Transformations for Ising networks}
%==============================================================================
The bond propagation algorithm can be applied to systems which possess
series, $Y$-$\Delta$, and $\Delta$-$Y$ transformations, including 2D Ising models, as suggested in Ref.~\onlinecite{lobb}.
%EC this was misleading -- sounded like lobb proposed delta-y for ising. 
%We now derive these ``building blocks" for 2D Ising models, an extension that was proposed 
%in Ref.~\onlinecite{lobb}.  
%YLL: reinstated ref to Lobb
We merely sketch a derivation of these known transformations for the Ising model,\cite{houtappel,syozi}  
with final forms optimized for computation.  
Consider the general Ising action
\begin{equation}
S(\{\sigma_i\},\{J_{ij}\}) = -\beta H =  \sum_{\left<ij\right> }J_{ij}\sigma_i \sigma_j
\label{eqn:rbim}
\end{equation}
where  the inverse temperature $\beta = {1 / k_B T}$, $H$ is the Hamiltonian, and the variables $\sigma = \pm 1$.
The nearest-neighbour {\em dimensionless couplings} $J_{ij}=\beta \widetilde{J}_{ij}$ are arbitrary real numbers.  
This has  rich physics: it includes the Edwards-Anderson spin glass model and the $\pm J$ random-bond Ising model (examples of disorder frustration), bond- and site-diluted Ising models (percolation physics), and the triangular Ising antiferromagnet (geometric frustration).

%%%%%%%% FIGURE:  Building Blocks %%%%%%%%%%%%%%%
\begin{figure}[Htb]
\psfrag{J12}{$J_{12}$}
\psfrag{J23}{$J_{23}$}
\psfrag{J31}{$J_{31}$}
\psfrag{J1}{$J_1$}
\psfrag{J2}{$J_2$}
\psfrag{J3}{$J_3$}
{\centering
\psfrag{Series}{(a) Series}
\subfigure
{\resizebox*{0.9\columnwidth}{!}{\includegraphics{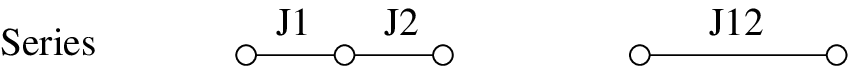}\label{fig:series}}}
\psfrag{Y-Delta}{(b) $Y$-$\Delta$}
\subfigure
{\resizebox*{0.9\columnwidth}{!}{\includegraphics{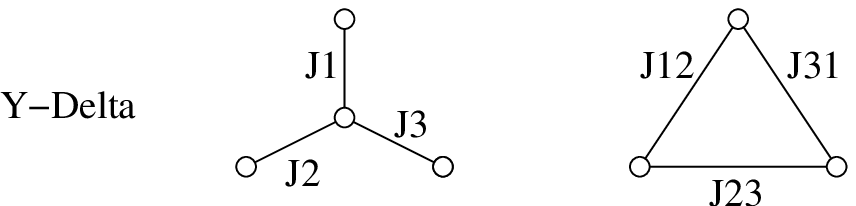}\label{fig:YDelta}}} \par}
\caption{Building blocks for the bond propagation algorithm.  
(a) {\em Series}.  In a series reduction, 
the middle spin is integrated out.  
(b) {\em $Y$-$\Delta$}. In a $Y$-$\Delta$ transformation, the middle spin is integrated out.
In the reverse ($\Delta$-$Y$) transformation, a spin is integrated back in.  
See the text for formulae relating the coupling constants in these transformations.
}
\label{fig:buildingblocks}
\end{figure}
%%%%%%%%%%%%%%%%%%%%%%%%%%%%%%%%%%%%%%

We define the building blocks as transformations of the $J$'s that preserve the value of the partition function,
$Z = \sum_{\{\sigma_i = \pm 1\}} e^{-\beta H}$.
A ``series" reduction corresponds to integrating out a spin with two neighbors, generating an effective coupling $j_{ij} = z_1{}^{1/2} z_0{}^{-1/2}$ between sites $i$ and $j$ as well as a constant `free energy' shift in the action $\delta f=z_1{}^{1/2} z_0{}^{1/2}$, where $z_0=\frac{1}{j_1j_2}+j_1j_2$ and $z_1=\frac{j_2}{j_1}+\frac{j_1}{j_2}$. 
(See Fig.~\ref{fig:series}.)
We have found it convenient to use the variables $j_i=e^{-J_i}$, $j_{ij}=e^{-J_{ij}}$, and $\delta f=e^{\delta F}$, because in this representation the transformations involve algebraic functions only (powers and roots) rather than transcendental functions and are thus more suitable for analysis and computation.

The $Y$-$\Delta$ transformation corresponds to integrating out the middle spin, $\sigma$,
in Fig.~\ref{fig:YDelta}.
Because of spin-flip symmetry, the only allowed terms in the effective action are {\em bilinear} in $\{\sigma\}$,
along with a constant free energy shift:
\begin{eqnarray}
Z_Y[\sigma_1,\sigma_2,\sigma_3] 
&=&\sum_{\sigma} e^{J_1\sigma\sigma_1+J_2\sigma\sigma_2+J_3\sigma\sigma_3} \nonumber \\
=Z_\Delta[\sigma_1,\sigma_2,\sigma_3] 
&=& e^{\delta F+J_{23}\sigma_2\sigma_3+J_{31}\sigma_3\sigma_1+J_{12}\sigma_1\sigma_2}~.
\end{eqnarray}
The couplings of the resulting ``$\Delta$" and the free energy shift are 
\begin{align}
j_{23} &= z_2{}^{1/4} z_3{}^{1/4} z_0{}^{-1/4} z_1{}^{-1/4} , \nonumber \\
\delta f &= z_2{}^{1/4} z_3{}^{1/4} z_0{}^{1/4} z_1{}^{1/4} ,
\label{eq:ydelta}
\end{align}
where $z_0 = \frac{1}{j_1j_2j_3} + j_1j_2j_3$, $z_1 = \frac{j_1}{j_2j_3} + 
\frac{j_2j_3}{j_1}$, and the expressions for $j_{31}$, $j_{12}$, $a_2$, $a_3$ are related to those above by cyclic permutations of the indices $1,2,3$.

\begin{figure}[Hbt]
{\centering
\subfigure[Specific heat $c(\beta)$]
{\resizebox*{0.9\columnwidth}{!}{\includegraphics{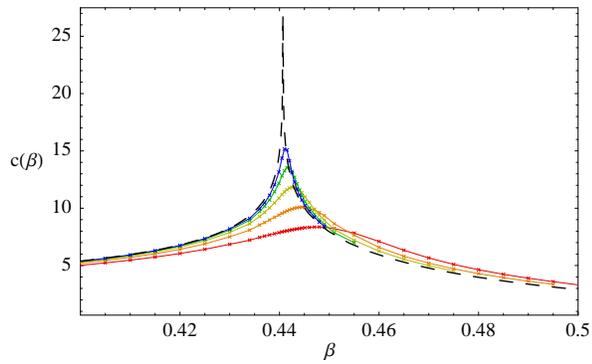}\label{fig:FSS}}} 
\subfigure[Scaled effective coupling $L^2 J_\text{eff} (\beta)$]
{\resizebox*{0.9\columnwidth}{!}{\includegraphics{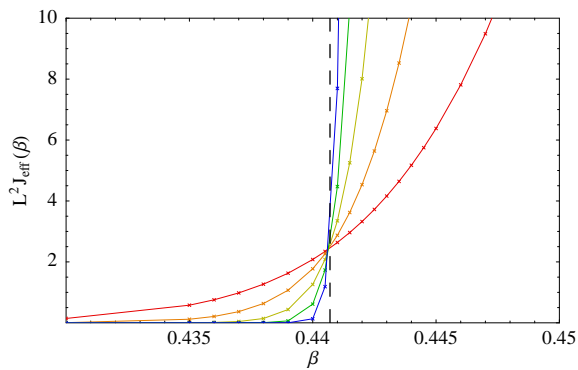}\label{fig:JeffFSS2}}} 
\par}
\caption{Results of bond propagation for the uniform ferromagnetic Ising model, for square lattices of side $L=64,128,256,512,1024$ with open boundary conditions.  
(a)  Specific heat $c(\beta)$ {\em vs.} inverse temperature $\beta$.  The black curve is the Onsager solution.
(b)  Effective corner-to-corner coupling scaled by system size, $L^2 J_{\rm eff}(\beta)$.  The crossing point indicates the transition temperature.}
\label{fig:proof}
\end{figure}

\begin{figure}[Htb]
{\centering
\subfigure[Specific heat $c(\beta)$]
{\resizebox*{0.9\columnwidth}{!}{\includegraphics{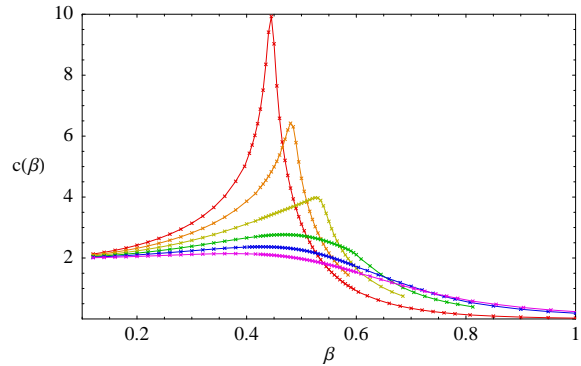}\label{fig:RBIMc}}} 
\subfigure[Effective coupling $J_\text{eff} (\beta)$]
{\resizebox*{0.9\columnwidth}{!}{\includegraphics{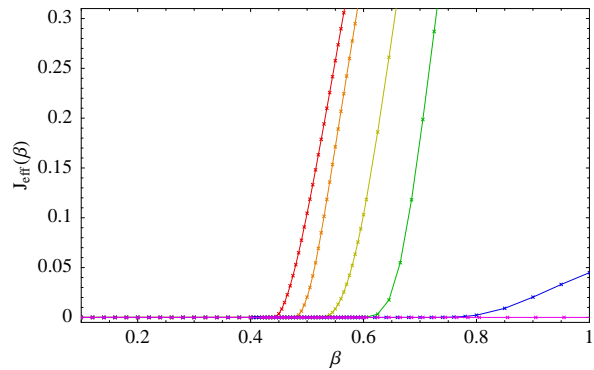}\label{fig:RBIMJeff}}}   
\par}
\caption{Results for the $\pm J$ RBIM on a $128 \times 128$ square lattice plotted as a function 
of the  inverse temperature, $\beta = 1/T$, for typical disorder configurations.
The concentrations of antiferromagnetic bonds are $p = 0, 0.025, 0.05, 0.075, 0.1,$  and $0.125$.  
(a) The peak in the specific heat broadens and shifts to lower temperature as $p$ is increased.
(b) The effective corner-to-corner coupling $J_{\rm eff}$ becomes nonzero in the ordered phase.}
\label{fig:RBIM}
\end{figure}

The inverse of the $Y$-$\Delta$ transformation is the $\Delta$-$Y$ transformation, which corresponds to integrating back in the middle spin, ``$\sigma$", of Fig.~\ref{fig:YDelta}.\cite{houtappel,syozi}  
The couplings of the resulting ``$Y$'' and the free energy shift are given by
\begin{align}
	j_1 &= \sqrt{\frac{1-t_1}{1+t_1}}, \nonumber \\
\delta f &= \frac{z_0}{j_1 j_2 j_3 + \tfrac{1}{j_1j_2j_3}},
\label{eq:deltay}
\end{align}
where 
$t_1 = c_2{}^{1/2} c_3{}^{1/2} c_0{}^{-1/2} c_1{}^{-1/2}$, with 
$c_0 = z_0 + z_1 + z_2 + z_3$ and  
$c_1 = z_0 + z_1 - z_2 - z_3$, and the $z_i$ are defined by
$z_0 = \tfrac{1}{j_{23} j_{31} j_{12}}$ and $z_1 = \tfrac{j_{31} j_{12}}{j_{23}}$,
and cyclic permutations.
%The forms of  $z_2,z_3$, etc., are defined by cyclic permutation.
These equations may be written in various forms that are much more efficient or robust in particular parameter regimes.  
For example, in Fig.~\ref{fig:proof} and in the $p=0$ case of Fig.~\ref{fig:RBIM}, we have used a formulation which is optimized for the uniform ferromagnetic case.
As emphasized in Ref.~\onlinecite{lobb}, infinite couplings (``shorts'') may appear during bond propagation, and need to be treated with care.

%==============================================================================
%%\mysection{Bond propagation for Ising networks}
%==============================================================================
%====================================================================
%\mysection{Computing Jeff}
%====================================================================
Now that we have the basic building blocks in place, the Frank-Lobb bond propagation
algorithm may be extended to 2D Ising models described by Eqn.~\ref{eqn:rbim}, as suggested in Ref.~\onlinecite{lobb}.
In its original form, the bond propagation algorithm
computes the effective resistance between two corners of a square network with open boundary conditions.  
When applied to Ising models on a square lattice,
it yields the ``renormalized" effective coupling 
$J_{\rm eff}$ between opposite corner spins.  
We can then trivially sum over the four configurations of these last two remaining spins to obtain the partition function of the original network, $Z(T)$.  
%EC removed -- we say it all later. 
%Since this is accurate to machine precision, it is possible to extract  other thermodynamic quantities such as the heat capacity $C(T)$ by, {\em e.g.}, numerical differentiation,\cite{whydifferencing}
% and see the emergent singularity at $T_c$ as system size is increased. 
%As shown below (see Fig.~\ref{fig:JeffFSS2}), 
%a more accurate determination of $T_c$ comes from analyzing $J_{\rm eff}$ directly,
%since it is indicative of the long range behavior of the correlation functions.  
%EC removed
%This quantity is indicative of the long-range behaviour of the correlation function 
%and should change rapidly near criticality (i.e., it should be sensitive to the bare coupling $J$ when this is near 
%the critical coupling $J_c=0.4407$).
%In this way, the effective coupling may be used as an indication of $T_c$.   

%====================================================================
%\mysection{FM Ising}
%====================================================================
%%%%%%%% FIGURE:  Ising Proof of Principle %%%%%%%%%%%%%%%
%%%%%%%%%%%%%%%%%%%%%%%%%%%%%%%%%%%%%%

As a proof of principle, we apply the algorithm to the uniform ferromagnetic Ising model with $\widetilde{J}_{ij}=+1$, and compare to the Onsager result for the infinite system\cite{onsager44}.
Fig.~\ref{fig:proof} shows the specific heat, $c(\beta) \equiv \frac{1}{L^2\beta^2} \frac{d^2 \ln Z}{d\beta^2}$,
estimated by second-order finite 
differencing\cite{whydifferencing} for various system sizes.  
A more natural diagnostic tool in this algorithm is the 
effective, ``renormalized" dimensionless coupling $J_{\rm eff}$ between spins on opposite corners of the original square lattice, which indicates whether long-range order is present.
The transition temperature may be accurately determined from the crossing 
point of $L^2 J_{\rm eff}$ plotted for various lattice sizes, as shown in
Fig.~\ref{fig:JeffFSS2}.

%==============================================================================
%\mysection{RBIM}
%==============================================================================
%%%%%%%% FIGURE:  RBIM Proof of Principle %%%%%%%%%%%%%%%
%%%%%%%%%%%%%%%%%%%%%%%%%%%%%%%%%%%%%% 

%YL:  small changes here,  2006-10-19
To illustrate that the method works for frustrated systems as well, we apply it to the $\pm J$ RBIM on a square lattice, where each ferromagnetic bond in the Ising model is replaced by an antiferromagnetic bond with probability $p$.
For this model it is known \cite{merz,wang2003} that the Curie temperature $T_c(p)$ decreases from
$T_c=2.2692$ at $p=0$ to $T_c=0.9533$ at $p_c=0.1093$.
Fig.~\ref{fig:RBIM} shows the results of bond propagation on typical disorder configurations for $128 \times 128$ lattices.  As the concentration of antiferromagnetic bonds $p$ is increased from $0$ to $0.125$, the peak in the specific heat $c(\beta)$ shrinks, changes shape and vanishes, indicating the destruction of the phase transition.  
The upturn in the effective coupling $J_{\rm eff} (\beta)$ (i.e., where $J_{\rm eff}$ begins to deviate from zero) is a useful indicator of $\beta_c=1/T_c$; the values thus obtained are in agreement with the phase diagram in Ref.~\onlinecite{merz}.

The presence of antiferromagnetic couplings introduces frustration.
According to Eq.~\ref{eq:deltay}, if the $\Delta$ couplings satisfy the inequality $\left( j_{31}{}^2 - j_{12}{}^2 \right) / \left( j_{31}{}^2 + j_{12}{}^2 \right) < j_{23}{}^2$, the $Y$ coupling $j_1$ is a {\em complex number}, due to the frustration  of the original $\Delta$.   
However, the partition function and effective coupling $J_{\rm eff}$ thus calculated remain real,
apart from small imaginary parts (of the order of $10^{-13}$) due to roundoff error.
%\textbf{
%However, we find that at the end of the calculation the effective coupling $J_\text{eff}$ and partition function $Z$ always come out real, apart from small imaginary parts (of the order of $10^{-13}$) due to roundoff error.
%}
%(In our current implementation for the $L=128$ system, $J_{\rm eff}$ remains real up to 
%$1$ part in $10^{-13}$ for the most frustrated case considered, $p=0.125$.)

%==============================================================================
%\mysection{TriAF}
%==============================================================================
The method can also address gaussian disorder and the case of full geometric frustration
(where every plaquette has an odd number of antiferromagnetic couplings), although errors accumulate faster 
in the frustrated case, and calculations are therefore reliable only for smaller system sizes or larger temperatures.
For example, the method is reliable for the 
fully geometrically frustrated case of a 
triangular antiferromagnet for temperatures above $0.25 \widetilde{J}$, $0.4\widetilde{J}$,  $0.7\widetilde{J}$, and $\widetilde{J}$ for $L = 4, 8, 16$, and $32$, respectively.  

%==============================================================================
%\mysection{Applicability}
%==============================================================================

Having shown that the algorithm works for unfrustrated systems as well as disordered/frustrated systems, we now discuss its range of applicability.
The bond propagation approach is applicable to all linear systems
\footnote{This category includes non-interacting tight-binding problems, semiclassical Heisenberg spins in the harmonic approximation, and resistor networks (for which bond propagation was originally formulated);  bond propagation can be understood as a linear algebra method for reducing certain pentadiagonal matrices.  In fact, bond propagation is a simpler alternative to nested dissection in certain situations; \emph{it may be possible to adapt bond propagation to Pfaffian elimination}, giving yet another route to the Ising partition function.
};
this work shows that it is also applicable to Ising models on planar graphs with no applied fields (in fact, bond propagation still works if fields are only present at the boundaries \cite{forthcoming}).
This includes models used for spin glasses, such as the $\pm J$ RBIM and the Edwards-Anderson model which chooses the couplings $J_{ij}$ from a gaussian distribution, and fully and partially frustrated Ising models.  
It does not include models which explicitly break $\mathcal{Z}_2$ symmetry in the  bulk, such as the Ising model in an applied field or the random field Ising model.  In this case,
$3$-spin couplings are allowed by symmetry upon equating the $Y$ and $\Delta$ partition functions,
and the resulting system of nonlinear equations for the coupling constants is overdetermined.  

\begin{figure}[Htb]
{\centering
\subfigure[]
{\resizebox*{0.7\columnwidth}{!}{\includegraphics{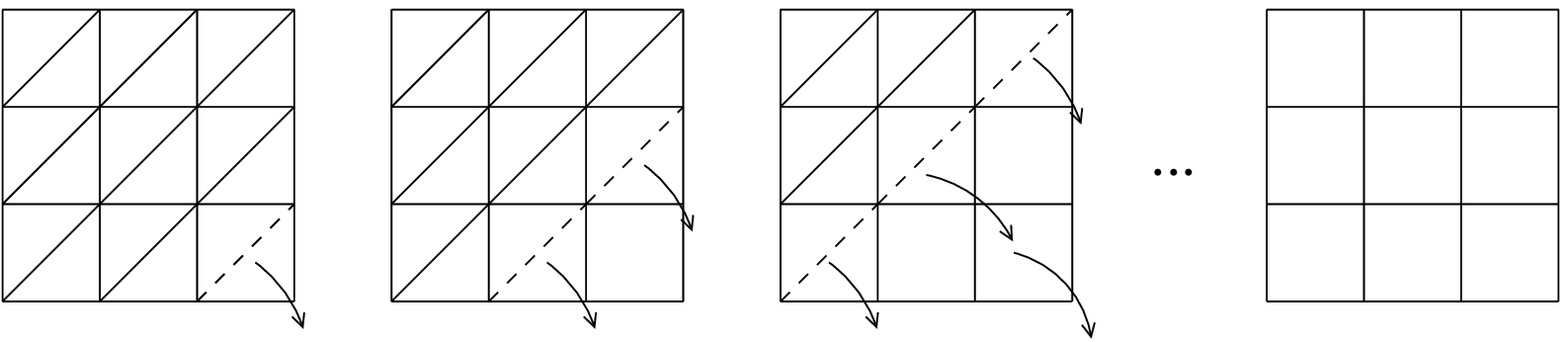}\label{fig:trilattreduction}}}
   \hspace{0.05\columnwidth}
\subfigure[]
{\resizebox*{0.2\columnwidth}{!}{\includegraphics{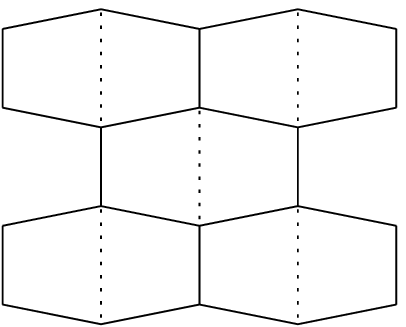}\label
{fig:honlattreduction}}}
\par}
\caption{(a) Reduction of a triangular lattice to a square lattice; (b) embedding of a honeycomb lattice in a square lattice using zero bonds (dotted lines).}
\label{fig:trihonreduction}
\end{figure}

%Rather extensive changes!
The method can be applied to any lattice which is a planar graph \footnote{That is, the lattice can be drawn in 
2 dimensions without any bonds crossing.  
%EC removed -- we lack space.  
%This corresponds to resistor network topologies that can be printed on a single-layer circuit board.
}, including square, triangular, honeycomb and kagome lattices, and even Bethe lattices and quasicrystals such as Penrose tilings.
Such lattices can be reduced to or embedded in a square lattice by propagating out ``effectively diagonal bonds'', by inserting zero bonds or infinite bonds, and/or by duality or $Y$-$\Delta$ transformations (see Fig.~\ref{fig:trihonreduction} for examples).
%This includes  square and triangular lattices,
%as well as honeycomb and kagome lattices, and even Bethe lattices and quasicrystals such as %Penrose tilings.   For a lattice with original coordination number $z>4$, some ``effectively %diagonal"
%bonds may need to be propagated out first, or  some nodes may need
%to be ``split" first (by turning one site into two, with an infinite coupling between them), in %order to reduce
%the coordination number to $z \le 4$.
The bond propagation algorithm requires open
boundary conditions in at least one direction in order to have a ``free edge" at which propagating
bonds can annihilate.  Therefore the algorithm can work with open boundary conditions in both directions, or cylindrical 
boundary conditions (open in one direction but periodic in the other), but not a torus.  
Cylinders with skew-periodic or helical boundary conditions may be used as well.
The bond propagation algorithm can also be straightforwardly adapted to infinite strips, as in Refs.~\onlinecite{merz}-\onlinecite{morgenstern1979}, and used to compute correlation lengths and free energy densities in that geometry.

%%%%%%%% FIGURE:  Triangular Lattice Reduction %%%%%%%%%%%%%%%
%%%%%%%%%%%%%%%%%%%%%%%%%%%%%%%%%%%%%%

%==============================================================================
%\mysection{pure maths}
%==============================================================================
Our algorithm is also interesting in a mathematical sense because
%, unlike the algorithms of Refs.~\onlinecite{saul1993} and~\onlinecite{merz}, 
it is not a generalization of one of the exact solutions of the uniform Ising model 
(such as the Onsager, Kaufman, or Kac-Ward solutions), 
nor  does it require fermion  or dimer mappings.
%dimer: KAsteleyn 1963
We believe that the existence of a $Y$-$\Delta$ equivalence for Ising systems, along with the fact that planar graphs are $Y$-$\Delta$/$\Delta$-$Y$-reducible, 
is a simple indicator of the ``hidden integrability'' of 2D zero-field Ising models which is responsible for the existence of seemingly unrelated exact solutions.  It is interesting to note that graph-theoretical methods have been used in a parallel body of work on zero-temperature RBIMs (e.g., Refs.~\onlinecite{bieche1980,boettcher2005}), and are restricted to 2D zero-field systems.

%==============================================================================
%\mysection{Conclusion}
%=============================================================================
In conclusion, we have developed an algorithm for solving 2D Ising models with arbitrary bond strengths on planar graphs.
The algorithm is a direct extension of the Frank-Lobb bond propagation algorithm for resistor networks\cite{lobb}.
It is able to reduce an Ising lattice completely using a sequence of local transformations,
thus allowing efficient, numerically exact computation of the partition function and correlation functions,
without relying on fermion or dimer mappings.   
The method requires negligible memory beyond the $O(L^2)$ required
to store the bond strengths,
and takes $O(L^3)$ time in general, and only $O(L^2 {\rm ln}L)$ for diluted models near percolation, for which it is the fastest method to our knowledge.
Parallelization is straightforward and can reduce
the required computational time to as little as $O(L)$.

It is a pleasure to thank R.~Fisch, E.~H.~Goins, C.~J.~Lobb, F.~Merz, and S.~A.~Kivelson for helpful discussions.  
This work was supported by Purdue University (YLL) and by the Purdue Research Foundation (EWC).  
EWC is a Cottrell Scholar of Research Corporation.

%How many lines can I add? 
%How many lines can I add? 
%How many lines can I add? 
%How many lines can I add? 
%How many lines can I add? 
%How many lines can I add? 
%How many lines can I add? 
%How many lines can I add? 
%How many lines can I add? 
%How many lines can I add? 
%How many lines can I add? 
%How many lines can I add? 
%How many lines can I add? 
%How many lines can I add? 
%How many lines can I add? 
%8 but not 9

\bibliographystyle{forprl}
\bibliography{rbim}

\begin{thebibliography}{10}
\bibitem{whydifferencing}
Eqns.~\ref{eq:ydelta} and \ref{eq:deltay} preserve the partition function $Z$
  but not its derivatives $\frac{d^nZ}{d\beta^n}$, hence the need for finite
  differencing. It is also possible to derive $Y$-$\Delta$ transformations
  which preserve derivatives of $Z$.\cite{forthcoming}.
\end{thebibliography}

%%%%%%%%%%%%%%%%%%%%%%%%%%%%%%%%%%%%%%%%%%%%%%%
\end{document}